\documentclass[reprint, superscriptaddress,amsmath,amssymb,aps,prl,longbibliography]{revtex4-2}

\usepackage{graphicx}
\usepackage{xcolor}
\usepackage{physics}
\usepackage{upgreek}
\usepackage{array}
\usepackage{braket}

\usepackage{multirow}

\usepackage{verbatim}

\usepackage[normalem]{ulem}
\usepackage[caption=false]{subfig}

\begin{document}
\title{A metastable state in $\mathrm{Yb}^+$ with lifetime greater than 10 minutes}
\author{Hassan Farhat}
\altaffiliation{These authors contributed equally to this work. Their emails are hassanfarhat@g.ucla.edu and williamgliu@g.ucla.edu.}
\author{William Liu}
\altaffiliation{These authors contributed equally to this work. Their emails are hassanfarhat@g.ucla.edu and williamgliu@g.ucla.edu.}
\author{Wesley C. Campbell}
\affiliation{Dept.\ of Physics and Astronomy, University of California Los Angeles, Los Angeles, California, USA}
\affiliation{Challenge Institute for Quantum Computation, University of California Los Angeles, Los Angeles, California, USA}
\affiliation{Center for Quantum Science and Engineering, University of California Los Angeles, Los Angeles, California, USA}

\date{\today}


\begin{abstract}
We find through direct, time-domain measurement of trapped $\mathrm{Yb}^+$ ions that the excited ${}^3[11/2]_{9/2}^o$ state has a collision-free lifetime of 17.8(6) min, long enough for use in atomic clockwork or quantum information processing.  We locate and benchmark a broad ``repump'' transition at $701\, \mathrm{nm}$ that readily transfers population back into the Doppler cooling cycle with the ground state, which may allow for quantum nondemolition measurement of sub-structure in the metastable manifold. Using a bi-stable Coulomb-crystal manometer, we find that the collisional quenching of this metastable ${}^3[11/2]_{9/2}^o$ manifold by background gas is more efficient than for ${}^2\mathrm{F}_{7/2}^o$, but nonetheless we observe an average lifetime exceeding 10 minutes at the base pressure of our apparatus.
\end{abstract}

\maketitle

Trapped ytterbium ions are flexible systems that have a wide range of applications. In even isotopes, the radiatively broadened linewidth of the ${}^2 \mathrm{S}_{1/2} \leftrightarrow {}^2 \mathrm{F}^{o}_{7/2}$ transition is of order nHz, and this transition is leveraged for optical frequency standards \cite{tofful_171yb_2024,Huntemann2012_E3} and the search for physics beyond the Standard Model \cite{hur_evidence_2022, Ono_Yb_Kingplot, yeh_2025}. For isotopes of ytterbium with spinful nuclei (${}^{171}\mathrm{Yb}^+$ and ${}^{173}\mathrm{Yb}^+$), the hyperfine sublevels within the ${}^2 \mathrm{F}_{7/2}^o$ state host metastable qubits that are optically separated from the ground state qubit, as demonstrated in the ``$omg$" scheme \cite{OMG}, and have been prepared with high fidelity \cite{Parks2026Magic}.

Recent proof-of-principle experiments with small-spin species ($J < 3$) have begun to demonstrate the use of nontrivial encodings within angular momentum manifolds for protecting coherent superpositions \cite{Gross2021Designing,Jain2024AbsorptionEmission,DeBry2026Error,Li2025Beating, Omanakuttan_Cat_Codes_2024}. Many of these schemes have been designed specifically to combat the errors that are most relevant for trapped-atom qubits, including Zeeman dephasing and spontaneous scattering, without the need to dedicate a large number of ions to protect a single qubit of quantum information. For these schemes, a larger array of errors can be rendered recoverable (i.e. the code distance can be increased) as the size of the manifold ($J$) is expanded, with some of the most-desirable codes requiring a minimum angular momenta of $J > 3$ for rank-1 absorption-emission (\AE) codes \cite{Aydin2025Class}, $J > 6$ for binary-octahedral spin codes \cite{Gross2021Designing}, and $J > 10$ for rank-2 \AE\ codes \cite{Aydin2025Class}. Additionally, in spin-cat encoding \cite{Omanakuttan_Cat_Codes_2024}, the suppression of logical errors scales exponentially with angular momenta. Though quantum error correction (QEC) has not yet been widely adopted for precision measurements, the basic idea that protecting coherence against environmental effects is desirable is a shared sentiment across many fields and it stands to reason that advances in QEC may pay dividends beyond quantum computing.  

The wealth of metastable, large-$J$ states contained within the electronic structure of ytterbium provides a valuable testbed for hosting qubits capable of realizing QEC codes. The multi-electron, odd-parity $J_1K$-coupled states with angular momenta $J\geq9/2$ are particularly useful in this respect, as decay pathways back to the Doppler cooling manifold and ${}^2 \mathrm{F}_{7/2}^o$ state are electric-dipole ($E1$) forbidden. 
The $J_1K$-coupled terms ${}^3[7/2]_{9/2}^o$,${}^1[11/2]_{9/2}^o$, and ${}^3[9/2]_{9/2}^o$ meet these criteria, and have recently been accessed via narrow electric quadrupole ($E2$) transitions from the ${}^2 \mathrm{F}_{7/2}^o$ state \cite{McMillin2026SubHertz}. However, because these excited states have magnetic dipole ($M1$) decay pathways to other lower-lying $J_1K$-coupled states, they may be more useful for entangling operations in the ${}^2 \mathrm{F}^o_{7/2}$ qubit than for long-term quantum information storage. In contrast, the state studied in this work (the ${}^3 [11/2]_{9/2}^o$ term at $30 \, 224 \, \mathrm{cm}^{-1}$ \cite{NIST_ASD}) lacks $M1$ decay pathways to other $J_1K$-coupled states, and meets the same criteria mentioned earlier. As a result, the ${}^3 [11/2]_{9/2}^o$ state has been identified a prime candidate for an additional qubit manifold \cite{McMillin2026SubHertz, Ackerman2026LongLived}.

In this Letter, we present laser spectroscopy of the ${}^2 \mathrm{F}_{7/2}^o \leftrightarrow {}^3[11/2]_{9/2}^o$ $E2$ transition in ${}^{171}\mathrm{Yb}^+$, ${}^{172}\mathrm{Yb}^+$ and ${}^{174}\mathrm{Yb}^+$. The collision-free lifetime of the ${}^3 [11/2]_{9/2}^o$ state is measured by recording the dwell time as a function of the reordering (``hopping") rate in a bi-stable three ion crystal, a method described in \cite{Hankin2019_Collisions}. Additionally, the ${}^3[11/2]_{9/2}^o \leftrightarrow {}^3[7/2]_{5/2}^o$ transition is demonstrated as an effective repump pathway for population recovery. Further, we provide measured hyperfine $A$ coefficients of the ${}^3[11/2]_{9/2}^o$ and ${}^3[7/2]_{5/2}^o$ states in ${}^{171}\mathrm{Yb}^+$.

We collect Doppler cooling fluorescence from individual $\mathrm{Yb}^+$ ions in a radio frequency Paul Trap with a magnetic field of $4.00(7)\,\mathrm{G}$ (applied to destabilize coherent dark states), using a photomultiplier tube (PMT), yielding tens of kilo-counts per second per ion, corresponding to a total collection efficiency of approximately 0.2$\%$. Removing an ion from the primary cooling cycle causes a discrete drop in fluorescence, enabling high-contrast discrimination between bright and dark ions.

To populate the metastable ${}^{2}\mathrm{F}^o_{7/2}$ state (the starting point for this spectroscopy), laser light drives the ${}^2\mathrm{S}_{1/2} \leftrightarrow \, {}^2\mathrm{D}_{5/2}$ $E2$ transition, which preferentially decays into the $^{2}\mathrm{F}^o_{7/2}$ state with a branching fraction of 0.83(3) \cite{tan_precision_2021}. The ${}^{2}\mathrm{F}^o_{7/2}$ population can be returned to the ground state via a laser-driven $E2$ transition to the $^1[3/2]^o_{3/2}$ state, which rapidly decays to the $^2\mathrm{S}_{1/2}$ ground state. Continuously applying light resonant with the $^2\mathrm{S}_{1/2} \leftrightarrow \, ^2\mathrm{D}_{5/2}$ transition and the $^2\mathrm{F}^o_{7/2} \leftrightarrow \, ^1[3/2]^o_{3/2}$ transition cycles population between $^2\mathrm{S}_{1/2}$ and $^{2}\mathrm{F}^o_{7/2}$, reducing the laser-induced fluorescence of the $^{2}\mathrm{S}_{1/2}^{} \leftrightarrow \, ^{2}\mathrm{P}^{o}_{1/2}$ cooling transition by roughly half.

Linear Coulomb crystals of up to six $\text{Yb}^+$ ions are prepared with spatial resolution to maximize signal discrimination. While cycling the ions between the $^2\mathrm{S}_{1/2}^{}$ and $^{2}\mathrm{F}^o_{7/2}$ states, light resonant with the $^2\mathrm{F}^o_{7/2} \leftrightarrow \, ^3[11/2]^o_{9/2}$ transition is applied. A successful transition of an ion into the ${}^3[11/2]^o_{9/2}$ state produces a sharp, quantized drop in PMT counts, which is verified on an EMCCD camera feed.  Using these quantum jump ``telegraph'' signals \cite{wlandQuantumjump}, we measure the $^2\mathrm{F}^o_{7/2} \leftrightarrow \, ^3[11/2]^o_{9/2}$ transition frequency in $^{172}\mathrm{Yb}^+$ and $^{174}\mathrm{Yb}^+$ to be $263.985\,96(2)\,\mathrm{THz}$ and $263.983\,92(2)\,\mathrm{THz}$ respectively. The $20\,\mathrm{MHz}$ uncertainties are primarily due to the incoherent population preparation in the Zeeman sublevels of the $^{2}\mathrm{F}^{o}_{7/2}$ state, which leads to Zeeman broadening. Following the same procedure described in \cite{McMillin2026SubHertz}, we measure the hyperfine $A$ coefficient of the ${}^3[11/2]^o_{9/2}$ state in $^{171}\mathrm{Yb}^+$ to be $A = 389(5)\,\mathrm{MHz}$.
 
The lifetime of the ${}^3[11/2]^o_{9/2}$ state is determined by observing quantum jumps back to the cooling cycle. Once the ${}^3[11/2]^o_{9/2}$ state is populated, light resonant with $^2\mathrm{S}_{1/2} \leftrightarrow \, ^2\mathrm{D}_{5/2}$ transition and $^2\mathrm{F}^o_{7/2} \leftrightarrow \,^3[11/2]^o_{9/2}$ transition are shuttered, while the Doppler cooling beam and repumping tones for the $^2\mathrm{D}_{3/2}$ and $^2\mathrm{F}^o_{7/2}$ states remain active to return non-shelved ions to the cooling cycle. We monitor the system for a sudden jump in fluorescence of an ion returning to the cooling transition. The dwell time measurements of population in the $^3[11/2]^o_{9/2}$ state have a resolution limited by the $100 \,\mathrm{ms}$ PMT binning time. It should be noted that the recorded dwell time also includes the duration of the $^{2}\mathrm{F}^{o}_{7/2} \rightarrow \, ^1[3/2]^o_{3/2}$ repump pathway; however, the $^{2}\mathrm{F}^{o}_{7/2}$ is depleted on a timescale significantly faster than the binning resolution. 

The collision-free radiative lifetime of the ${}^3[11/2]^o_{9/2}$ state was measured through extrapolation by performing dwell time measurements while varying the pressure of the room-temperature vacuum chamber. The steady-state pressure inside the vacuum chamber was controlled by passing a constant current through a non-evaporative getter (NEG), and the  pressure was monitored with an Agilent IPCMini pump controller. 

\begin{figure}
  \centering
  \subfloat[]{%
    \label{fig:zig}%
    \includegraphics[width=0.35\linewidth, height=2cm]{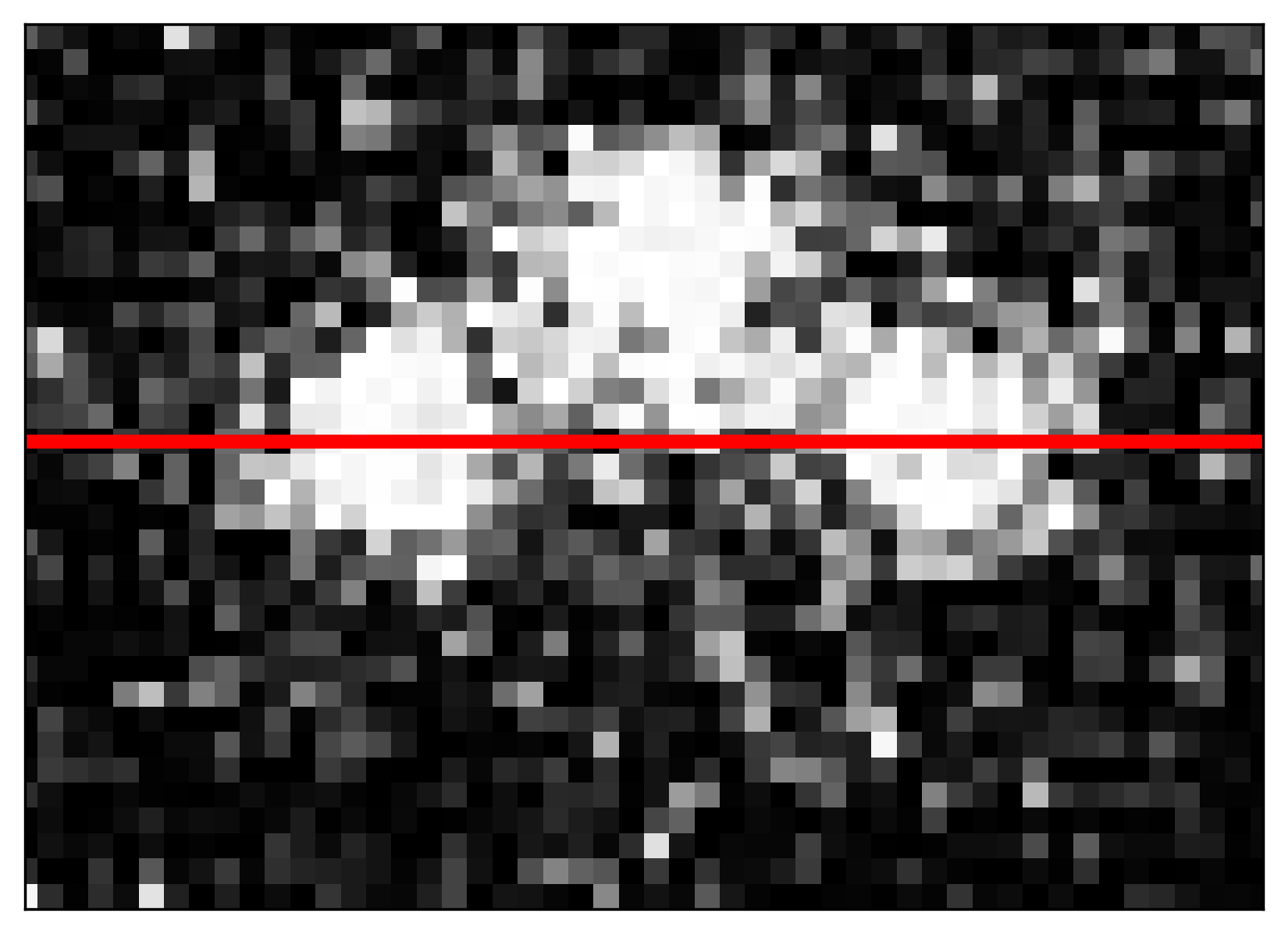}%
  }%
  \quad
  \subfloat[]{%
    \label{fig:zag}%
    \includegraphics[width=0.35\linewidth, height=2cm]{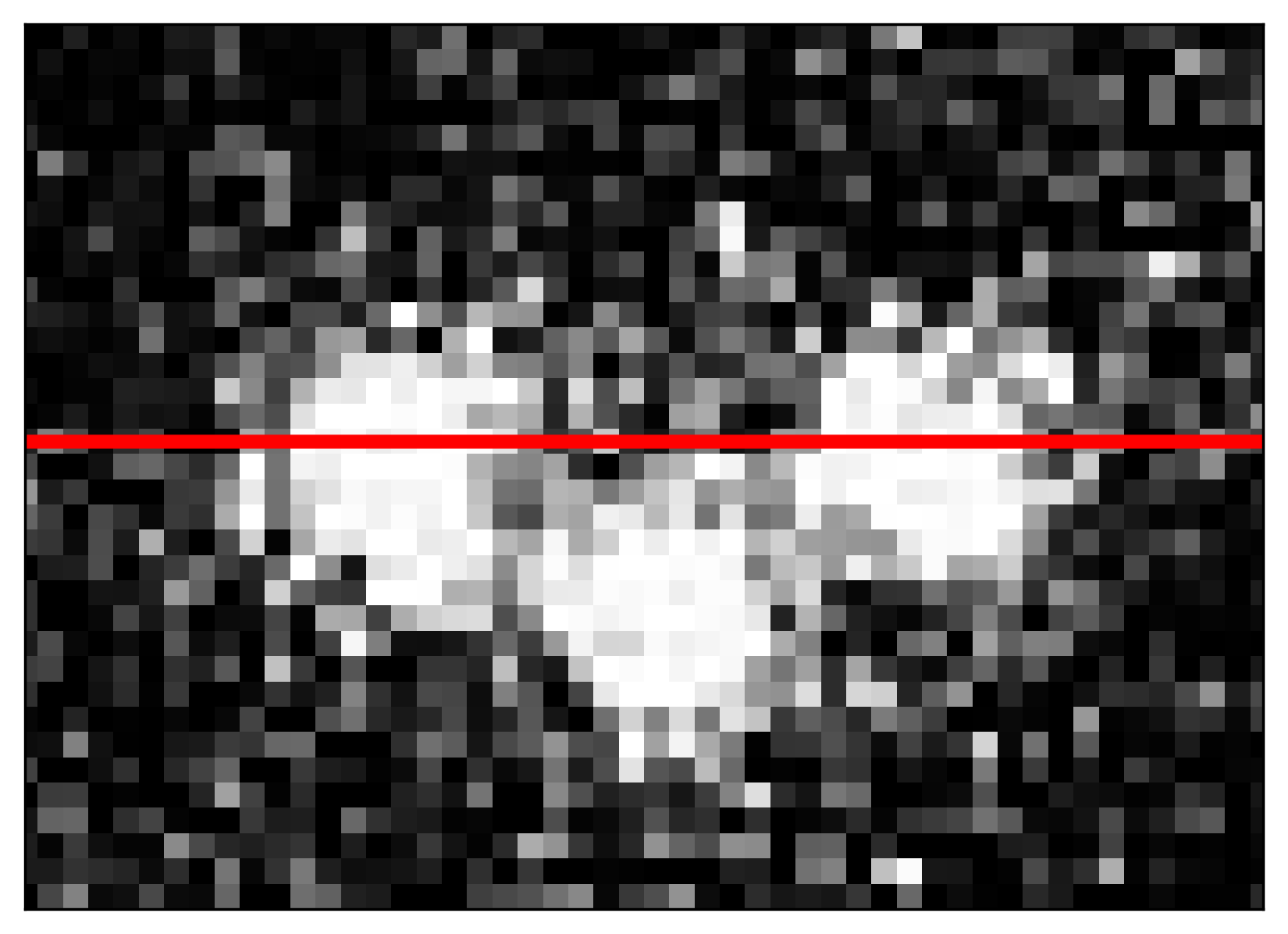}%
  }\\[-2pt]
  \subfloat[]{%
    \label{fig:hist}%
    \hspace*{-27pt}%
    \includegraphics[width=0.9\linewidth]{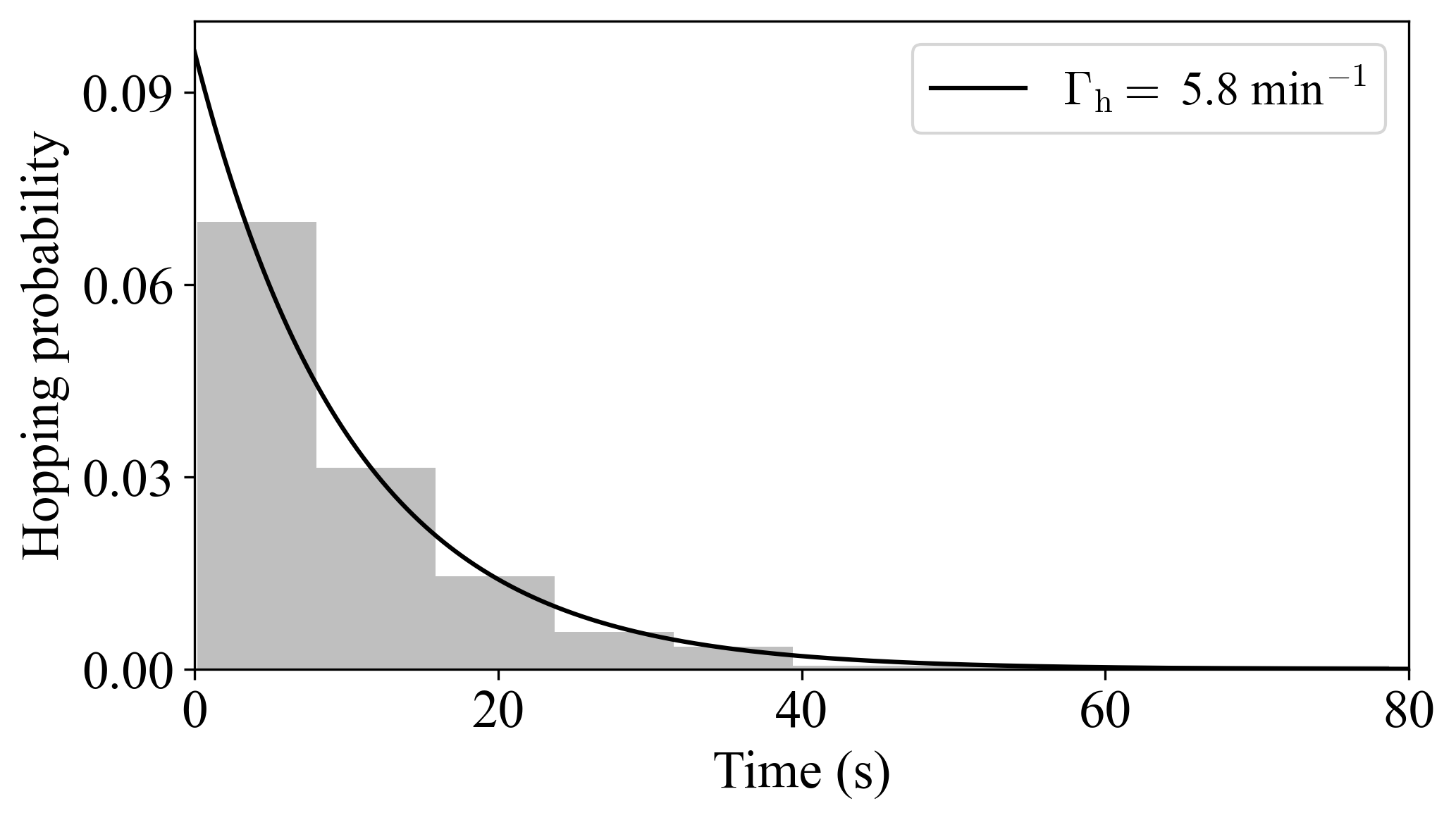}%
  }\\[-6pt]
  \caption{Upper: ECCDM frames showing a kinked chain of ${}^{174}\mathrm{Yb}^+$ ions in the ``zig" (\ref{fig:zig}) and ``zag" (\ref{fig:zag}) modes. Lower: The average hopping rate $\Gamma_{\mathrm{h}}$ for a given dataset was determined by dividing the total number of hops by the total observation time.}
  \label{fig:zigzag}
\end{figure}

\begin{figure}
    \includegraphics[scale=0.5]{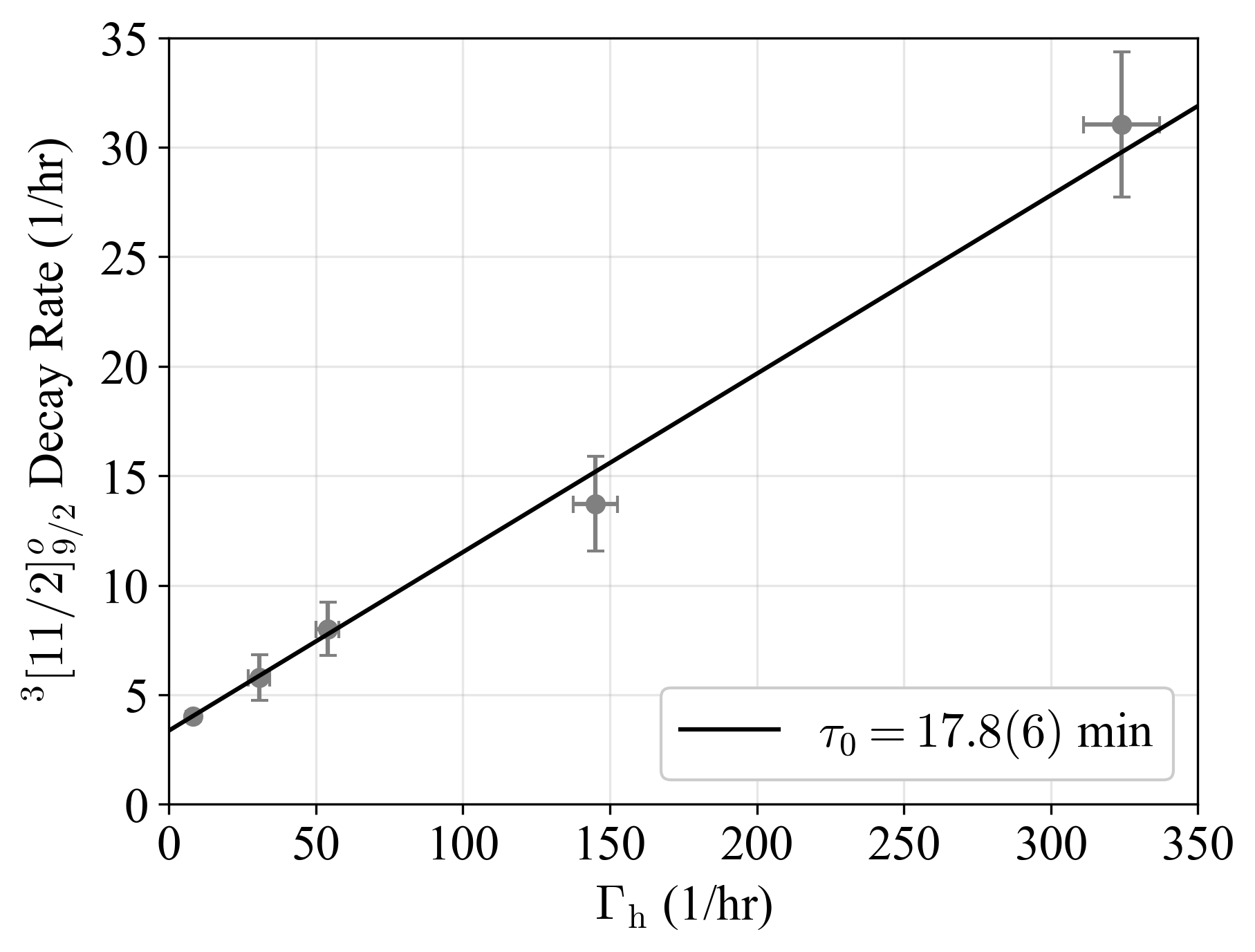}
    \caption{Observed dwell time in the ${}^3 [11/2]_{9/2}^{o}$ state as a function of $\Gamma_{\mathrm{h}}$, the average hopping rate of the center ion in the kinked, three-ion crystal shown in Fig. \ref{fig:zigzag}. The collision free lifetime $\tau_0$ is taken to be the reciprocal of the observed dwell time at $\Gamma_{\mathrm{h}} \rightarrow 0$.}
    \label{fig:unlimited_pressure} 
\end{figure}

To avoid contaminating the lifetime measurements with the limited accuracy of the ion pump gauge in the ultra-high vacuum regime, we measured the local collision rate in a three-ion crystal at various nominal pressures, providing an accurate measure of \emph{relative} pressure with a large \emph{absolute} calibration uncertainty. The three ion crystal was loaded and converted into a kinked, ``zig-zag'' configuration by elevating the trap endcap voltage, shown Fig. \ref{fig:zigzag}. At this endcap voltage setting, the central ion can ``hop" between two discrete stable positions: above (``zig") or below (``zag") the crystal axis. The EMCCD camera continuously imaged the crystal over large periods for collision-induced hops between the two configurations. The frequency of these discrete mode hops provides a precise measurement of how the local pressure at the crystal changes. The measured lifetimes are plotted against the average crystal hopping rate at each pressure reading in Fig. \ref{fig:unlimited_pressure}. The collision-free lifetime of the ${}^3[11/2]^o_{9/2}$ state is found through extrapolation, yielding $\tau = 17.8(6)\,\mathrm{minutes}$, which is congruent with a recent theoretical calculation of $40.93\,\mathrm{minutes}$ \cite{Ackerman2026LongLived}. Due to selection rules, the decay rate is assumed to be dominated by the ${}^3[11/2]_{9/2}^o \rightarrow {}^2 \mathrm{F}^{o}_{7/2}$ $E2$ decay pathway. As a result, the lifetime-limited linewidth of the ${}^2 \mathrm{F}^o_{7/2} \leftrightarrow {}^3[11/2]_{9/2}^o$ transition is $1/\tau \approx 2\pi \times 200 \,\upmu\mathrm{Hz}.$

\begin{figure}
    \includegraphics[scale=0.86]{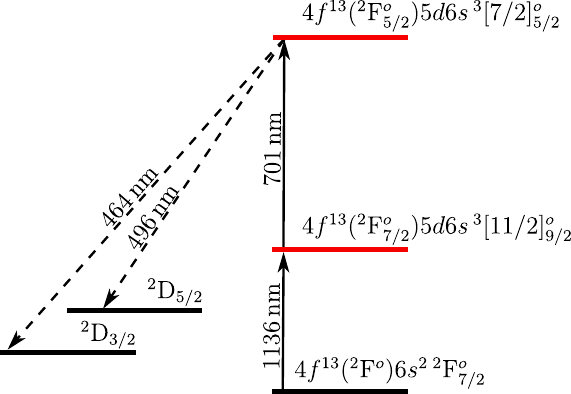}
    \caption{Partial level structure of $\mathrm{Yb}^+$. Laser excitation is shown with solid lines and dashed lines are electric dipole decays. The ${}^3[11/2]^o_{9/2}$ state is where population is shelved and the ${}^3[7/2]^o_{5/2}$ state is used as a repump to move population back to the cooling cycle.}
    \label{fig:grot}
\end{figure}

A repump pathway is useful to facilitate fast readout of a clock based on this transition or a hyperfine qubit defined on the ${}^3[11/2]^o_{9/2}$ state due to its long lifetime. Similar to the $^2\mathrm{F}^o_{7/2} \leftrightarrow \, ^1[3/2]^o_{3/2}$ repump transition, we investigated $E2$ transitions to $J_{1}K$-coupled states that have favorable laser frequencies and $E1$ decay channels back to the cooling cycle. The  ${}^3[11/2]^o_{9/2}\leftrightarrow{}^3[7/2]^o_{5/2}$ transition at $701 \, \mathrm{nm}$, as shown in Fig. \ref{fig:grot}, was selected because the excited state (with electron configuration $4f^{13}(^2\mathrm{F}^o_{5/2})5d6s$ and energy level $44 \, 497.51 \, \mathrm{cm}^{-1}$) has $E1$ decay channels to the $^2\mathrm{D}_{3/2}$ and $^2\mathrm{D}_{5/2}$ states, which are already effectively repumped in our experiment. Additionally,  the wavelength $701 \, \mathrm{nm}$ is within the search range of our Ti:sapphire laser.

Spectroscopy of the ${}^3[7/2]^o_{5/2}$ state was performed by first preparing a chain of multiple $\mathrm{Yb}^{+}$ ions in the ${}^3[11/2]^o_{9/2}$ state. The cooling light and repumping tones for the $^2\mathrm{D}_{3/2}$ and the $^2\mathrm{F}^o_{7/2}$ states are left active while the driving laser is shuttered. The Ti:Sapphire laser frequency was then scanned until multiple ions returned to the cooling cycle at a significantly faster rate than the average lifetime of the ${}^3[11/2]^o_{9/2}$ state. This process was repeated until we were able to consistently repump multiple ions simultaneously. We measured the ${}^3[11/2]^o_{9/2} \leftrightarrow \, ^3[7/2]^o_{5/2}$ transition frequency in $^{172}\mathrm{Yb}^+$ and $^{174}\mathrm{Yb}^+$ to be $427.89878(2) \,\mathrm{THz}$ and $427.89886(2) \,\mathrm{THz}$ respectively.

\begin{figure}
    \includegraphics[scale=0.86]{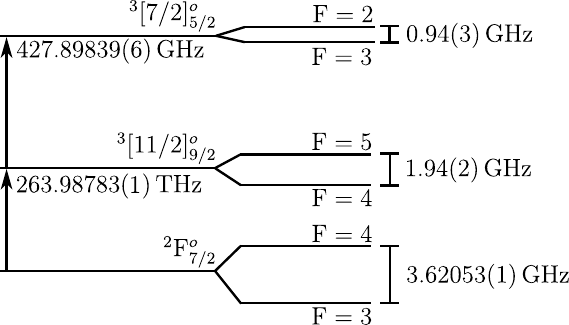}
    \caption{{Hyperfine structure of the ${}^3[11/2]_{9/2}^o$ and ${}^3[7/2]^o_{5/2}$ states. By selectively populating one of the hyperfine sublevels in ${}^2 \mathrm{F}^o_{7/2}$, we determines the ${}^2 \mathrm{F}^o_{7/2} \leftrightarrow {}^3[11/2]_{9/2}^o$ and ${}^3[11/2]^o_{9/2}\leftrightarrow  {}^3[7/2]^o_{5/2}$ transition centroid frequencies depicted above.}}
    \label{fig:hyperfine_struct}
\end{figure}

In addition, spectroscopy of the hyperfine levels of the ${}^3[7/2]^o_{5/2}$ state in $^{171}\mathrm{Yb}^+$ was performed. The search was simplified by first populating the $F=5$ manifold of the ${}^3[11/2]^o_{9/2}$ state via the $\vert{}{}^2\mathrm{F}^o_{7/2}\, , F=3 \rangle \leftrightarrow \vert{}{}^3[11/2]^o_{9/2} \, F=5 \rangle$ transition. Due to $E2$ selection rules, the only allowed transition to the ${}^3[7/2]^o_{5/2}$ manifold from this state is $\vert{}{}^3[11/2]^o_{9/2}, F = 5 \rangle \leftrightarrow \vert{}{}^3[7/2]^o_{5/2}, F = 3 \rangle$. After driving this transition, we subsequently drove the $\vert{}{}^3[11/2]^o_{9/2}, F = 4 \rangle\leftrightarrow \vert{}{}^3[7/2]^o_{5/2}, F = 2\rangle$ transition to determine the hyperfine $A$ coefficient to be $A=-310(10)\,\mathrm{MHz}$, as presented in Fig. \ref{fig:hyperfine_struct}.

We have demonstrated that it may be possible to operate a state-of-the-art hyperfine  ${}^3[11/2]_{9/2}^o$ qubit in ${}^{171}\mathrm{Yb}^+$ with relative ease, given its 17.6(8) minute collision-free lifetime and its readily accessible radio-frequency hyperfine splitting of $1.94(2) \,\mathrm{GHz}$. The additional metastable manifold augments the current existing $omg$ scheme in ${}^{171}\mathrm{Yb}^+$; by defining the $g$- and $m$- type qubits on the ${}^2 \mathrm{F}_{7/2}^o$ and ${}^3[11/2]_{9/2}^o$ manifolds respectively, both qubit manifolds are optically isolated from the Doppler cooling cycle during active operation. In addition, the ${}^3[11/2]_{9/2}^o$ manifold may be utilized as a logical qubit manifold in even isotopes in spin-cat encoding, in which the computational states are defined on the stretch states $m_J=\pm 9/2$ \cite{Omanakuttan_Cat_Codes_2024}.

\medskip
\begin{acknowledgments}
\textit{Acknowledgments ---} The authors acknowledge Nils Huntemann for helpful discussions.  This work was supported by ARO W911NF-24-S-0004, NSF PHY-2207985 and OMA-2016245, and the Gordon and Betty Moore Foundation DOI: 10.37807/GBMF11566.   
\end{acknowledgments}

\bibliography{Yb1136Lifetime}

\end{document}